\documentclass{llncs}
\RequirePackage{filecontents}
\RequirePackage{comgipp}

\newcommand{\theReferenceText}{\fullcite{Schubotz2019}}
\usepackage[
	backend=biber,
	style=numeric,
	giveninits=true,
	url=false,
	isbn=false,
	]{biblatex}
\usepackage{amsmath}

\usepackage{booktabs} %
\usepackage{multirow}
\usepackage{listings}
\usepackage[breaklinks]{hyperref}
\usepackage{cleveref}
\ifdefined\theReferenceText
{}
\else
\usepackage[moderate]{savetrees}
\fi
\usepackage{graphicx}

\newcommand{\zbl}[1]{\href{https://zbmath.org/?q=an:#1}{\textsf{#1}}}
\newcommand{\mlt}[1]{ \begin{tabular}[c]{@{}l@{}}#1\end{tabular}}
\newcommand{\queryLink}{\href{https://zbmath.org/?q=py\%3A2007-2018+\%26+\%28+ab\%3A\%22editorial+remark\%22+\%7C+ab\%3A\%22editorial+note\%22+\%29+\%26+\%28+\%22very+similar\%22+\%7C+\%22high+similarity\%22+\%7C+overlap+\%7C++plagiari*+\%7C+identical+\%7C+substantial*+\%7C+essentially+\%29+\%21\%28+so\%3Aieee+\%7C+se\%3A00000250+\%7C+se\%3A00001661+\%7C+pu\%3AAIP+\%7C+an\%3A0584.10010+\%7C+an\%3A0712.35001+\%7C+an\%3A0597.14041+\%7C+an\%3A1375.14126+\%7C+an\%3A0156.05104+\%7C+an\%3A1345.15011+\%7C+an\%3A1262.11083+\%7C+an\%3A1360.47003\%29}{Click here to execute the query. Subscription required for full display!}}
\begin{document}
\mainmatter

\title{Forms~of~Plagiarism in Digital Mathematical Libraries}

\author{Moritz Schubotz\inst{1,2} \and Olaf Teschke\inst{2} \and  Vincent Stange \inst{1}  \and Norman Meuschke\inst{1} \and Bela Gipp\inst{1}}

\institute{University of Wuppertal, Wuppertal, Germany\\
  \email{last@uni-wuppertal.de}
  \and FIZ Karlsruhe / zbMATH, Berlin, Germany\\
  \email{olaf.teschke@fiz-karlsruhe.de}}

\maketitle
\ifdefined\theReferenceText
\thispagestyle{firststyle}
\fi

\begin{abstract}
We report on an exploratory analysis of the forms of plagiarism observable in mathematical publications, which we identified by investigating editorial notes from zbMATH.
While most cases we encountered were simple copies of earlier work, we also identified several forms of disguised plagiarism.
We investigated 11 cases in detail and evaluate how current plagiarism detection systems perform in identifying these cases.
Moreover, we describe the steps required to discover these and potentially undiscovered cases in the future.
\end{abstract}

\section{Introduction}\label{sec.intro}
Plagiarism is `the use of ideas, concepts, words, or structures without appropriately acknowledging the source to benefit in a setting where originality is expected'~\cite{Fishman09,Gipp2014}. Plagiarism represents severe research misconduct and has strongly negative impacts on academia and the public. Plagiarized research papers compromise the scientific process and the mechanisms for tracing and correcting results. If researchers expand or revise earlier findings in subsequent research, papers that plagiarized the original paper remain unaffected. Wrong findings can spread and affect later research or practical applications.

Furthermore, academic plagiarism causes a significant waste of resources \cite{Foltynek2019}. Reviewing plagiarized research papers and grant applications causes unnecessary work. For example, Wager \cite{Wager2014} quotes a journal editor stating that 10\% of the papers submitted to the respective journal suffered from plagiarism of an unacceptable extent. If plagiarism remains undiscovered, funding agencies may even award grants for plagiarized ideas or accept plagiarized research papers as the outcomes of research projects. Studies showed that some plagiarized papers are cited at least as often as the original \cite{Long2009}. This is problematic, since publication and citation counts are widely used as indicators of research performance, e.g., for funding or hiring decision. Moreover, universities and research institutions invest considerable resources to investigate and sanction plagiarism.

The waste of resources and the deterioration of academic quality due to plagiarism is also a pressing concern for zbMATH. The zbMATH abstracting and reviewing service organizes and reviews the world's literature in mathematics and related areas since 1931. zbMATH includes its predecessor {\it Jahrbuch f\"ur die Fortschritte der Mathematik}, which goes back to 1868. As of today, zbMATH comprises about 4 million publications, reviewed with the help of more than 7'000 international domain experts. These expert reviewers do not repeat the foregone peer-review process of the venue publishing the article in question. Rather the zbMATH reviewers provide an unbiased view on the originality and innovative potential of articles for their subject areas, i.e., across venues. Ensuring the quality of articles is an integral part of zbMATH's mission statement and central to the usefulness of the service.

However, so far zbMATH's reviewing process for articles is conducted manually without the help of automated originality checks. Quality and originality insurance relies entirely on the knowledge of the expert reviewers who have to identify potential content overlap with other papers. The ability to spot such similarities requires reviewers to be closely familiar with similar articles. If an article exhibits significant overlap with prior works without discussing this fact, the reviewers will assign an editorial remark to the article.

The growth in publications indexed by zbMATH has raised concerns regarding the viability of the entirely manual reviewing process. Thus, zbMATH is investigating the use of automated approaches to support their reviewers. To guide and support this process, we performed an exploratory analysis of cases for which zbMATH reviewers questioned the originality of articles or identified plagiarism with certainty. Our goal is to develop an understanding of the typical characteristics of mathematical articles with questionable originality to derive the requirements for an automated system that can detect such cases.

We structure our report on this investigation as follows.
\Cref{sec.background} discusses related work before \Cref{sec.approach} presents the methodology of our investigation. \Cref{sec.res} describes our findings and discusses their implications. \Cref{sec.concloutl} summarizes our results and gives an outlook on the next steps of the project.

\section{Background and Related Work}\label{sec.background}
The problem of academic plagiarism has been present for centuries \cite{Weber-Wulff2014}. However, the advancement of information technology has made plagiarizing easier than ever \cite{Foltynek2019}. Forms of academic plagiarism range from copying content (\textit{copy\&paste}) over ``patch-writing", i.e., interweaving text from multiple sources with moderate adjustments, to heavily concealing content reuse, e.g., by paraphrasing or translating text, and reusing data or ideas without proper attribution \cite{Weber-Wulff2014}. The easily recognizable copy\&paste-type plagiarism is more prevalent among students whose main motivation for plagiarizing is typically to save time \cite{McCabe05}. Concealed forms of plagiarism are more characteristic of researchers, who have strong incentives to avoid detection \cite{Alzahrani2012}.

While making plagiarizing easier, information technology also facilitated the detection of plagiarism. Researchers proposed many plagiarism detection approaches that employ lexical, semantic, syntactical, or cross-lingual text analysis \cite{Meuschke13,Eisa2015,Gupta2016}. Such approaches typically employ a two-stage process consisting of candidate retrieval and detailed analysis \cite{Stein2007a,Meuschke13}. In the candidate retrieval stage, the approaches employ computationally efficient retrieval methods to limit the collection to a set of documents that may have been the source for the content in the input document. In the detailed analysis stage, the systems perform computationally more demanding analysis steps to substantiate the suspicion and to align components in the input document and potential source documents that are similar \cite{Meuschke13, Alzahrani2012, Foltynek2019}.

Current plagiarism detection approaches reliably detect copied or moderately altered text; some approaches are also effective for finding paraphrased and translated text. Most plagiarism detection systems available for productive use focus on reliably and efficiently identifying plagiarism forms with little to no obfuscation, which are characteristic of students. We conjecture that student plagiarism is a more profitable market segment for commercial providers of plagiarism detection services. Arguments in favor of this hypothesis are the higher number of students compared to researchers, the higher frequency of plagiarism among students than among researchers \cite{Swazey93,McCabe05} and the availability of well-established, efficient methods to find literal text reuse \cite{Meuschke13,Eisa2015,Gupta2016}.

The market leader for plagiarism detection services, \textit{iParadigms LLC}\footnote{https://www.crunchbase.com/organization/iparadigms-inc}, offers its products under several brand names. The system \textit{turnitin}\footnote{https://www.turnitin.com} is tailored to providing originality checks and academic writing training for students. The \textit{iThenicate}\footnote{https://www.turnitin.com/products/ithenticate} service is mainly offered to academic publishers and conference organizers for checking research publications. The iThenticate service is also licensed to other academic service providers, such as \textit{Crossref} who offers the service to its members as \textit{Similarity Check}\footnote{https://www.crossref.org/services/similarity-check/}.

The detection methods employed by commercial systems, such as turnitin and iThenticate, are trade secrets. However, the performance of the systems in benchmark evaluations \cite{HTW-PDS} suggests that they mainly use efficient text retrieval methods, such as word-based fingerprinting and vector space models. Text fingerprinting approaches first split a document into (possibly overlapping) word or character $n$-grams, which are used to create a representation of the document or passage (the ‘fingerprint’) \cite{Meuschke13}. To enable efficient retrieval, most approaches select a subset of the fingerprints, which they store in an index. To speed up the comparison of fingerprints, some approaches hash or compress the fingerprints, which reduces the lengths of the strings to compare and allows for computationally more efficient numerical comparisons \cite{Foltynek2019}.

To complement the many text analysis approaches and to improve the detection capabilities of concealed forms of academic plagiarism, researchers proposed approaches that analyze nontextual content features, such as academic citations \cite{Gipp11,Gipp11c,Gipp2014,gipp14a,Gipp13b,Meuschke14,Pertile2016}
and images \cite{Meuschke2018}. Nontextual content features in academic documents are a valuable source of semantic information that are largely independent of natural language text. Considering these sources of semantic information for similarity analysis raises the effort plagiarists must invest for obfuscating reused content \cite{Meuschke14,Meuschke2019}.

Nontextual feature analysis appears to be a promising approach to plagiarism detection for mathematics and related fields. In these fields, much of the semantic content of publications is expressed in terms of mathematical notation. Research showed that classical text retrieval methods, which are similar to the methods employed by commercial plagiarism detection services, are less effective for documents in mathematics, physics and other domains that routinely interweave natural language and mathematical notation \cite{Wolska08}.

However, only a few studies have addressed the detection of plagiarism in digital mathematical libraries \cite{Meuschke2017a,Meuschke2018a,Meuschke2019} regardless of the detection approach. We briefly describe the main findings of these studies hereafter.

Following up on discussions at the doctoral consortium of the \mbox{SIGIR} conference 2015,
\citeauthor{Meuschke2017a} (2017) described mathematics-based plagiarism detection (MathPD) as a discrete sub-problem within mathematical information retrieval. The authors argued that the different approaches to query formulation and query processing distinguish MathPD from the mathematical document retrieval problem as defined by \citeauthor{Guidi2016} \cite{Guidi2016}.

To test whether an exclusive analysis of mathematical similarity holds promise for plagiarism detection, \citeauthor{Meuschke2017a} (2017) gathered documents that have been retracted for plagiarism and contain significant amounts of mathematical content from three sources. The first source was an earlier study on retracted publications by \citeauthor{Halevi2016} \cite{Halevi2016}. This study had queried Elsevier's full text database \textit{Science Direct} for the term "RETRACTED" in October 2014. The search yielded 988 retracted articles, of which 276 had been retracted for plagiarism \cite{Halevi2016}. \citeauthor{Meuschke2017a} limited these 276 publications to a set of 39 publications that contain significant amounts of mathematical content. The second source was the blog \textit{Retraction Watch}\footnote{\url{http://www.retractionwatch.com}} from which \citeauthor{Meuschke2017a} obtained two confirmed cases of plagiarism. The third source was the crowd-sourced project \textit{VroniPlag}\footnote{\url{http://www.vroniplag.wikia.com}}, from which \citeauthor{Meuschke2017a} obtained three additional cases. The authors then limited the 44 cases they had gathered to cases that matched their area of expertise, i.e., mathematics, physics, and computer science, which resulted in 19 cases the authors then investigated manually. They categorized the types of shared mathematical content they observed in the analysis of these 19 cases into six broad categories \cite{Meuschke2017a}:
\begin{description}
\item [Identical:] an exact copy of math in the source document.
\item [Equivalent:] equivalent forms, e.g., due to commutativity or distributivity.
\item [Order changes:] order of expressions within document differs.
\item [Different presentation:] structurally and semantically identical.
\item [Splits or merges:] of expressions that are semantically identical.
\item [Different concepts:] different, yet semantically (nearly) identical, concepts, e.g., summation over vector components instead of matrix multiplication.
\end{description}

Of the 19 cases reviewed manually, \citeauthor{Meuschke2017a} selected 10 cases that were most representative of the types of content similarity they observed. The authors converted the ten source documents and ten retracted documents of those cases from PDF to LaTex using InftyReader \cite{Suzuki2004} and subsequently from LaTeX to XHMTL using LaTexML\footnote{\url{http://dlmf.nist.gov/LaTeXML/}}. They embedded the converted documents in the dataset of the NTCIR-11 Math task~\cite{Aizawa2014} (105'120 arXiv documents). Using pairwise document comparisons, \citeauthor{Meuschke2017a} evaluated the retrieval effectiveness of similarity measures that consider basic representational math features, i.e., identifiers, numbers, operators, and combinations thereof. The best performing approach, a set-based comparison of the frequency of mathematical identifiers, retrieved eight of ten test cases at the top rank and achieved a mean reciprocal rank of $0.86$.

In a follow-up study, \citeauthor{Meuschke2019} (2019) introduced similarity measures that consider the order of mathematical identifiers and presented a two-stage retrieval process consisting of a candidate retrieval and a detailed analysis stage that replaced the exclusive use of pairwise document comparisons \cite{Meuschke2019}. They implemented the process in the HyPlag prototype that also offers a user interface to investigate the identified similarities \cite{Meuschke2018a}. The candidate retrieval stage employs efficient index-based retrieval methods based on mathematical features. The candidate documents retrieved in the first stage then undergo pairwise comparisons in the detailed analysis stage of the process. The authors compared the effectiveness of their math-based analysis to citation-based and text-based approaches. They found that the order-observing similarity measures for mathematical identifiers achieved better results than the order-agnostic measures in their previous study \cite{Meuschke2017a}. Most of their ten test cases also exhibit a high textual similarity. A combined analysis of math-based and citation-based similarity performed equally well as the text-based analysis for the ten test cases. In an analysis of all 105K documents in their dataset, \citeauthor{Meuschke2019} demonstrated that the combined analysis of math-based and citation-based similarity has advantages over a text-based analysis by identifying interesting instances of content reuse that a text-based analysis could not detect.

In summary, \citeauthor{Meuschke2019} (2019) demonstrated that the combined analysis of text-based and nontextual similarity, e.g., the similarity of mathematical content and citations, achieves promising results for retrieving confirmed cases of plagiarism that involve mathematical content. However, the set of documents they analyzed is small and likely does not reflect the full spectrum of possible content reuse in digital mathematical libraries. To aid in the advancement of plagiarism detection methods for mathematics and related disciplines, we analyze a larger set of cases obtained from the zbMATH collection. An interesting property of this collection is that it exclusively includes published work since 2007, i.e., from a period in which larger publishers already employed services like CrossRef's Similarity Check. Also digitization has reached maturity during this period, so we can expect that these examples were not detected by standard tools.

\section{Method}\label{sec.approach}
Subsection \ref{sc.reusefilter} describes our approach to identify cases of \emph{noticeable content reuse} (NCR).
Subsection \ref{sc.Ppf2Tei} presents the challenges we faced in regard to processing the cases using the HyPlag detection system.
Lastly, Subsection \ref{sc.smallScale} defines the properties we evaluate for a small number of example cases.

\subsection{Identification of noticeable content reuse}\label{sc.reusefilter}
The zbMATH collection currently contains 3'981'836 publications. Since 2007, zbMATH follows a intra-organizational procedure for marking publications that exhibit noticeable content reuse. The managing editor and the deputy editor-in-chief decide on a case-by-case basis on the actions to be taken if illegitimate content reuse is observed.
From 2007 to 2018, 1'226'203 new publications have been added to the zbMATH database.
Although all publications underwent peer-review by the publishing venue before being submitted to zbMATH, 446 cases of questionable content reuse have been reported to the managing editor of zbMATH. After careful investigation of these reports, 149 cases received an editorial note about NCR. This list includes cases of content reuse by the same authors in different papers as well as content reuse by different authors. Moreover, in contrast to \citeauthor{Meuschke2019} (2019), this dataset includes both true positives (149), i.e., suspicious documents, and false positives (302).

The list of the 149 suspicious cases is openly available from the zbMATH website\footnote{\url{https://zbmath.org}} using the query shown in Listing \ref{lst.zbl}. The logic of the query is as follows: Line 1 filters for publications that appeared between 2007 and 2018, which matches the period investigated in this paper. `py' is the query term for publication year.
Line 2 filters the abstracts (ab) for the keywords `editorial remark' or `editorial note.'
Line 3 filters for remarks that indicate plagiarism. Examples for excluded editorial notes are incorrect results, conceptual flaws, or organizational comments. Note that we applied this filter to all fields, not just the abstract, since sometimes the indicator for an editorial note is included in other fields, e.g., the keywords.
For example the, document \zbl{1191.35223}\footnote{We use zbMATH identifiers for referring to cases throughout the paper. The identifiers resolve via, e.g., \url{https://zbmath.org/1191.35223}  to documents accessible without subscription.} is in the category `suspected plagiarism,' although this is not indicated in the abstract of the document.
Line 4 corrects the search result and excludes false positives.\footnote{See \url{https://zbmath.org/general-help/} for the details of the search syntax.}

\begin{lstlisting}[breaklines=true, breakatwhitespace=true, label=lst.zbl, numbers=left,float=t,caption=zbMATH search query to retrieve papers with a noticeable amount of content reuse. \queryLink ]
py:2007-2018 &
( ab:"editorial remark" | ab:"editorial note" ) &
( "very similar" | "high similarity" | overlap |  plagiari* | identical | substantial* | essentially )
!( so:ieee | se:00000250 | se:00001661 | pu:AIP | an:0584.10010 | an:0712.35001 | an:0597.14041 | an:1375.14126 | an:0156.05104 | an:1345.15011 | an:1262.11083 | an:1360.47003)
\end{lstlisting}

\subsection{The challenge of full text availability}\label{sc.Ppf2Tei}

After having identified the set of cases to investigate, we had to import the documents to our plagiarism detection system HyPlag (cf. Section \ref{sec.background}), which requires text encoding initiative (TEI) format as input. Unfortunately, the zbMATH dataset does not include full texts. The full texts might be obtained by following the DOI, an arXiv link, or a link to another digital repository. Yet, in most cases (except for arXiv) the LaTeX sources of the documents are unavailable. While HyPlag includes a conversion procedure that generates the TEI input format from PDF files via the PDF processor GROBID\footnote{\url{https://grobid.readthedocs.io}}, this procedure cannot process mathematical formulae. As a result, TEI documents generated from PDF documents miss mathematical formulae.

In a previous study \cite{Meuschke2017a}, we employed the image to formula conversion tool InftyReader (cf. \Cref{sec.background}). However, the results were unsatisfactory. While InftyReader extracted parts of the formulae correctly, its extraction accuracy regarding the structure of the document was significantly worse than that of GROBID. We also evaluated several alternative tools, including the combination of maxtract \cite{maxtract} and pdfminer\footnote{\url{https://pypi.org/project/pdfminer/}}. %
We also discovered and tested new machine learning based approaches\footnote{for example: \url{http://cs231n.stanford.edu/reports/2017/pdfs/815.pdf}}. However, these approaches only worked in a few exceptional cases.

Eventually, we were unable to identify a tool capable of converting the zbMATH documents including all essential features, i.e, text, formulae and figures. Consequently, we modified the setup of our study and decided to investigate only a small sample of the collected documents. To still analyze a diverse and representative test dataset, we decided to use a semi-random method. The managing editor of zbMATH manually selected interesting cases that had undergone extended internal discussions, which we see as an indicator that the potential overlap was not initially obvious. Due to the diversity of the cases and the different decisions zbMATH reviewers made regarding the legitimacy of the observed similarities, the documents can form an interesting test collection for PD systems in mathematics. The small test collection consists of the 11 items listed in Table \ref{tb.overview}.

\subsection{Properties to investigate}\label{sc.smallScale}

After choosing the test cases, we specified the following properties to investigate:
\begin{enumerate}
    \item Which type of plagiarism is observable in the article?
    \item Is the article digitally available? Has it been retracted?
    \item How important were the text, the figures, the references, and the formulae during the discussion about issuing an editorial remark?
    \item When were the articles published? What are the languages of the documents?
    \item What is the impact of the case? Has the article that received an editorial remark been cited. Does other statistical data on the use of this article exist?
    \item Which requirements on a plagiarism detection system derive from  this case?
\end{enumerate}
\section{Results} \label{sec.res}
In this section, we first present descriptive statistics for the cases exhibiting noticeable content reuse. Second, we describe the 11 selected test cases in detail.
Third, we derive requirements on a plagiarism detection system from the exemplary results of our analysis.

\subsection{Distribution of the source documents}\label{sec.statistics}
\begin{figure}
\centering
    \includegraphics[trim=2.5cm 19.5cm 8.5cm 2.5cm,clip,width=.48\textwidth,]{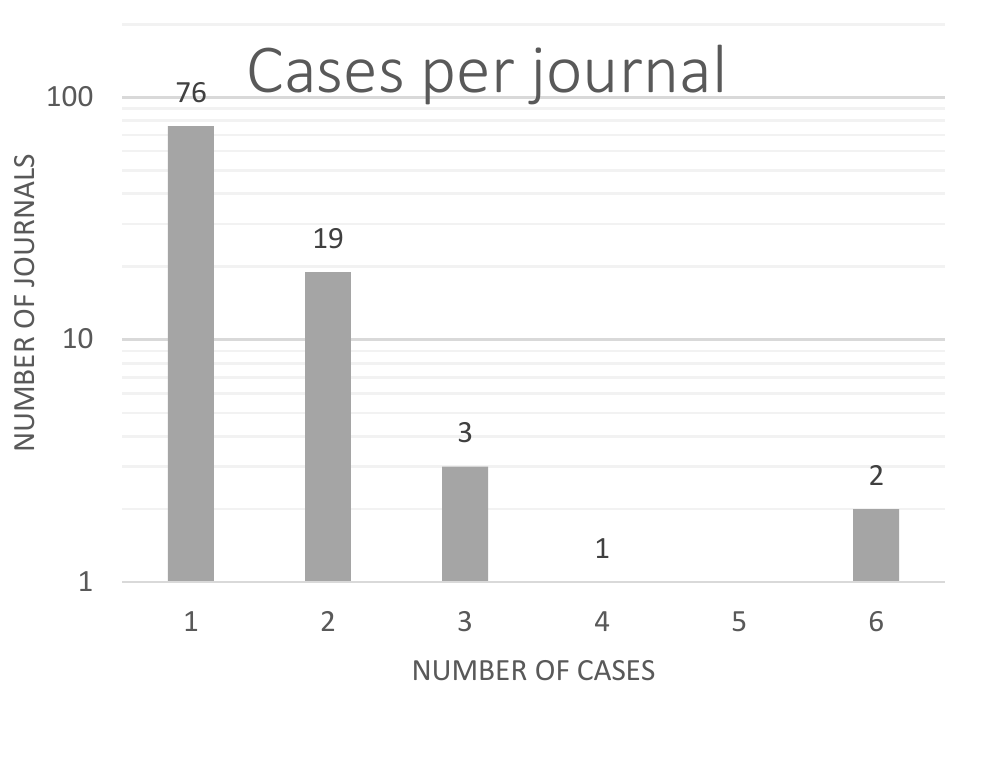}
    \includegraphics[trim=2.5cm 19.5cm 8.5cm 2.5cm,clip,width=.48\textwidth]{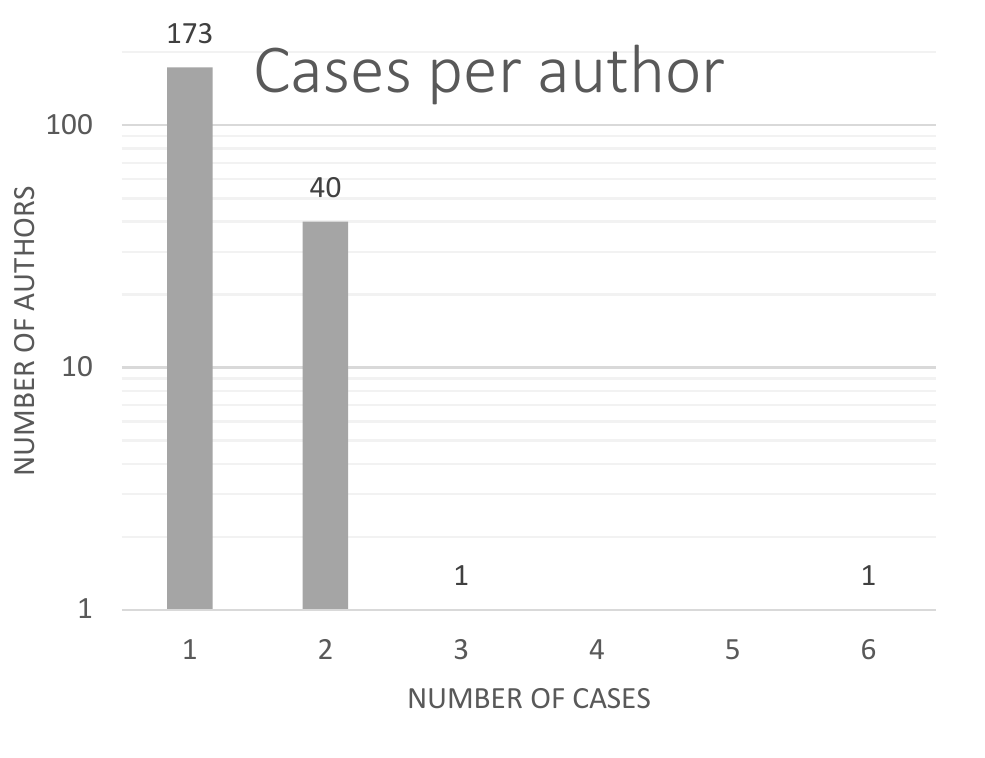}
    \caption{Distribution of journal (left) and author (right) frequencies on a logarithmic scale.}
    \label{fig:srcDist}
\end{figure}

The 149 publications that received an editorial remark for NCR consist of 139 journal articles and 5 books.
The 139 articles originate from 101 journals. Two journals published 6 articles with editorial remarks each (cf. Figure \ref{fig:srcDist}), while 76 journals published only one such article. The 149 documents have 215 authors. One person was an author of 6 publications that received an editorial remark, while 173 authors were involved in only one case (cf. Figure \ref{fig:srcDist}). For 78 of the 149 documents, the full texts are available from at least one source (DOI (76), arXiv (3), eudml (4), or emis (3)). See Appendix~\ref{sec.list} for the complete list of cases.
\begin{table}[tp]
\caption{Overview of manually investigated cases. The similarity scores were computed using HyPlag. $l_i$ denotes the later document of case $i$ and $e_{i,j}$ the $j$-th earlier document of case $i$.}
\label{tb.overview}
\begin{tabular}{@{}lllllrrp{3cm}@{}}
\toprule
{\#} & $l_i$          & $e_{i,j}$                                  & \textbf{retr.} & \textbf{\mlt{avail. \\ via}}     & \textbf{\mlt{text\\sim}} & \textbf{\mlt{cit\\sim}} & \textbf{spread}  \\ \midrule
1 & \zbl{1349.46021} (2015) & \mlt{\zbl{06696052} (2015) \\ \zbl{1353.46015} (2014)} & - & \mlt{doi,\\ eudml,\\ ams}  & \textbf{.23}    & -    &  40 reads on researchgate \\
2 & \zbl{1381.51005} (2008) & \zbl{1162.51304} (2007) & yes & \mlt{ams}         & .03           & -             & -            \\
3 & \zbl{1119.11307} (2001) & \zbl{1062.11019} (2000) & yes & \mlt{doi,\\ ams}   & \textbf{.33}  & -             & 40 reads on researchgate, 21 downloads on springer link \\
4 & \zbl{1112.35034} (2005) & \zbl{0632.65108} (1987) & - & \mlt{doi,\\ ams}     & \textbf{.33}  & -             & 6 reads on researchgate \\
5 & \zbl{1183.05037} (2008) & \zbl{0247.05143} (1972) & - & \mlt{doi,\\ ams}     & .0            & \textbf{.75}  & 5 reads on researchgate \\
6 & \zbl{1121.35118} (2005) & \zbl{1062.81046} (2004) & - & \mlt{doi,\\eudml,\\arXiv}            & .06           & -             & 5 reads \& 3 cits. on researchgate, 61 downloads on springer link \\
7 & \zbl{1176.08001} (2009) & \zbl{1036.08001} (2003) & - & -            & -           & -             & - \\
8 & \zbl{1219.30004} (2011) & \zbl{1040.30002} (2004) & - & \mlt{doi,\\eudml}     & .14           & .19           & 21 reads \& 4 cits. on researchgate \\
9 & \zbl{0946.35085} (1999) & \zbl{0816.47056} (1994) & - & \mlt{doi}     & .09           & -             & 4 reads \& 1 cit. on researchgate, 27 downloads on springer link \\
10 & \zbl{1155.35429} (2006) & \zbl{1142.35593} (2004) & - & \mlt{doi}    & .16          & -              & 112 reads \& 41 cit. on researchgate, 38 cits. on ScienceDirect \\
11 & \zbl{1360.81259} (2017) & \mlt{\zbl{1185.81005} (2009) \\ \zbl{--} (-) } & - & \mlt{doi}       & 0.1                     & -               &  -            \\ \bottomrule
\end{tabular}
\end{table}

\subsection{Manual investigation}
Hereafter, we present our findings from manually analyzing the 11 test cases:
\paragraph{Case 1 (legitimate content reuse)}
consists of the inspected document $l_1$ and the earlier works $e_{1,1}$ and $e_{1,2}$.

While the zbMATH editorial remark just reads: ``Almost the same results were previously obtained in ($e_{1,1}$)", more information is available through the reviews at \url{https://mathscinet.ams.org/mathscinet-getitem?mr=3390281}:
\begin{itemize}
    \item Theorem 2.3(2) is Theorem 2.4(7)
    \item Corollary 2.4 is Corollary 2.6
    \item Corollary 2.5(2),(4) and (5) is Corollary 2.5(5),(3) and (7)
    \item Theorem 2.7 and its Corollary 2.8 are respectively Theorem 2.11 and Corollary 2.12 of the older document (by noting from Theorem 2.11 of the older document that the order bounded operators $T:E\to E$ in Corollary 2.12 may be replaced by positive ones).
\end{itemize}
Moreover, the reviewer states for the second prior document $e_{1,2}$
\begin{itemize}
    \item Note also that Theorem 2.3(5) and Corollary 2.5(8) are respectively Theorem 2.7(8) and Corollary 2.10(5) of the second prior document.
    The proofs of the results that are republished in the paper under review are similar to those of the other publication.
\end{itemize}

\begin{figure}[tb]
\noindent\includegraphics[width=\textwidth]{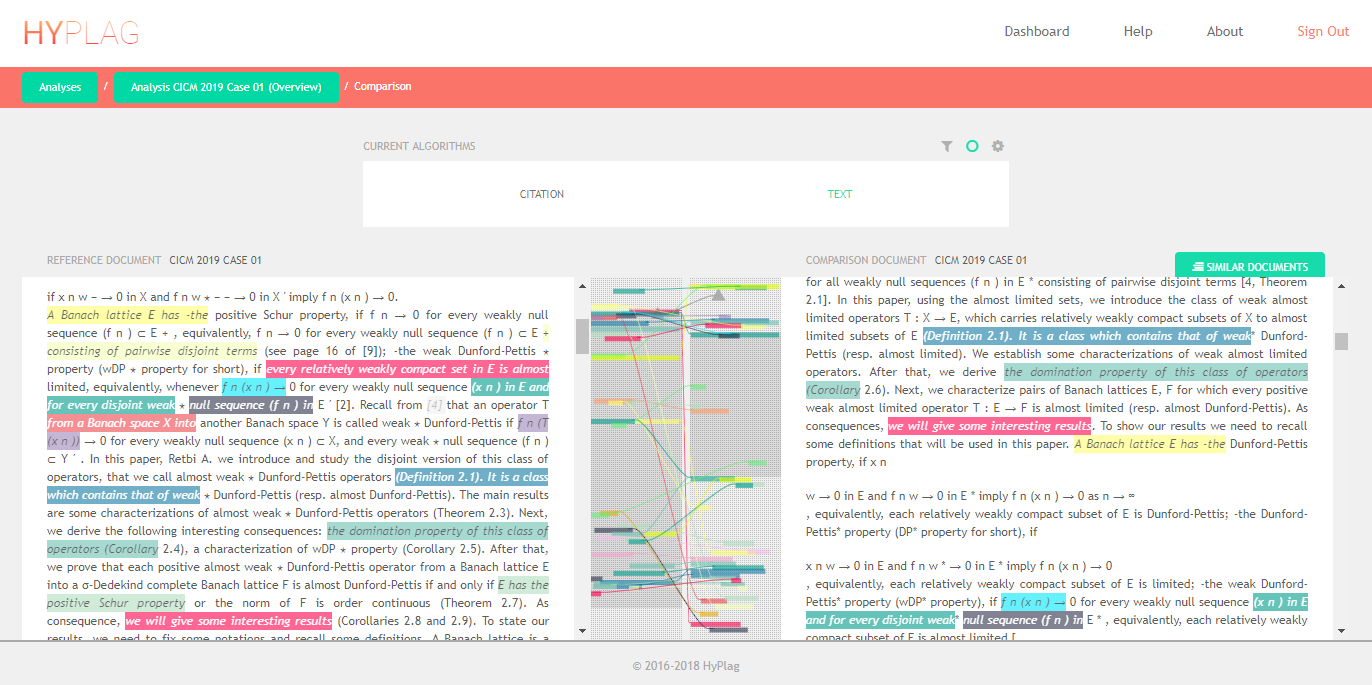}
\caption{Visualization of case 1 in HyPlag prototype.}\label{fg.case1}
\end{figure}

Despite the overlap in content, the editor decided to publish the article. The rationale of this decision is that the similar structure seems to be the obvious solution to the given problem. However, this explanation does not justify the high textual similarity (0.23) that HyPlag computed for $l_1$ and $e_{1,1}$. Despite this moderately high value\footnote{A value above 0.2 is considered as suspicious.} for the overall textual similarity, the HyPlag visualization\footnote{\url{https://www.hyplag.org/} user cicm@hyplag.org pw: cicm2019} (cf. Figure \ref{fg.case1}) shows that the text similarity in the section described by the reviewers is particularly high. To our knowledge, neither the reviewer, nor the editor had access to a text similarity visualization such as the one of HyPlag when they decided not to retract the article.

Note that \zbl{1394.47040} also used theorems from $e_{1,1}$ in combination with results from \zbl{1336.46019}. The authors published an erratum \zbl{06644537} and acknowledged that `some results' have been proven by $e_{1,1}$ and \zbl{1336.46019}. They `feel sorry' that they have ignored the original sources. In contrast to the other sources \zbl{1394.47040} draws upon, We did not identify a textual similarity of \zbl{1394.47040} with $e_{1,1}$ and \zbl{1336.46019}.

\paragraph{Case 2 (retracted)} consists of the retracted paper $l_2$ which overlaps with $e_2$. The retraction note states `The author and the editor regret an oversight that this paper has significant overlap with the prior publication.'

From a plagiarism detection perspective, this case is particularly interesting since the later paper does not reuse text from the earlier paper (text similarity score 0.03). Moreover, the documents do not have common citations. However, many formulae and figures in the documents can be used to identify the overlap.

Unfortunately, the extraction of the figures from the PDF was challenging. State of the art image extraction technology was unable to identify the images. The internal representation of the figures in the PDF object stream prevents the identification of the basic geometric objects as a meaningful figure.

\paragraph{Case 3 (retracted)} consists of the paper $l_3$ which is similar to $e_3$ without citing the earlier work. The later paper reuses several identical formulae and has a significant text overlap (text similarity score 0.33). The paper was published in 2001 and retracted in 2007. In the meantime, it received 2 citations in zbMATH - interestingly, by two papers that also cited $e_3$.

\begin{figure}[tb]
\includegraphics[width=\textwidth]{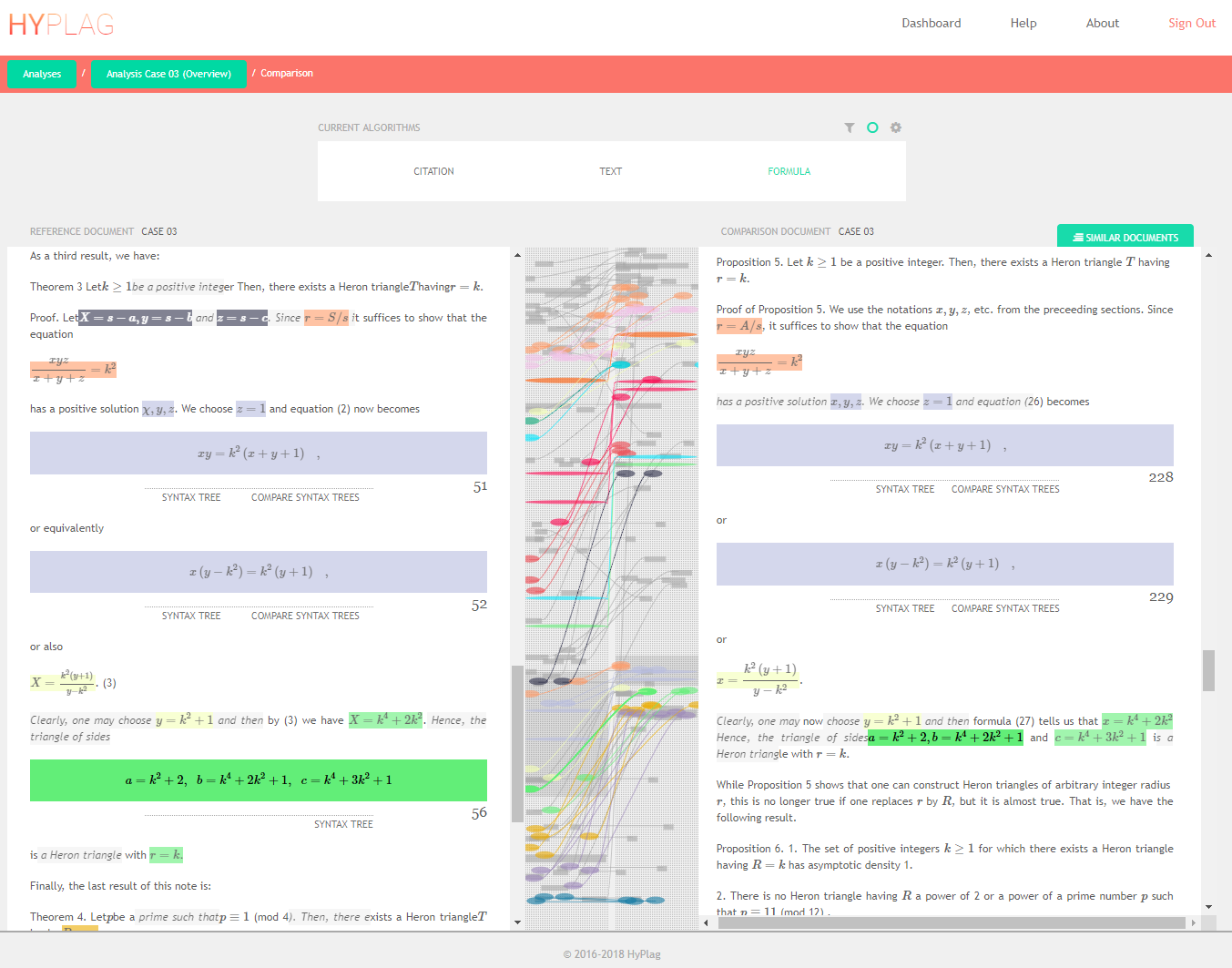}
\caption{Visualization of case 3 in HyPlag prototype.}\label{fg.case3}
\end{figure}

For this case, the LaTeX sources were available to us. Thus, we can visualize the formula similarity using HyPlag. Despite the high text similarity, the identical formulae are apparent within the visualization (cf. \Cref{fg.case3}).
\paragraph{Case 4 (plagiarism)} In this case, the authors of \zbl{1253.35026}, which they also published as $l_4$, copied results from $e_4$ published in 1987. The papers have a high text similarity of 0.33, even though we extracted the text from the earlier paper using OCR. However, one may speculate that the limited availability of the original in digital form (only available as a scan) contributed to the late discovery of the case. The republished version of the article $l_4$ is unavailable in digital form. \zbl{1253.35026} has been cited once. The case illustrates the limitations resulting from insufficient  digitisation.

\paragraph{Case 5 (translation)} is a prototypical example for translation plagiarism. The original work $e_5$ written in French and published 1972 was translated to $l_5$ published in 2008 without acknowledging the original source.

This case likewise illustrates the limitations induced by sub-optimal digitisation. While both papers are available via DOI, the scan of the original only allows for an OCR quality that is insufficient for matching the citations in both documents. For instance, using Adobe Acrobat version 11.0.0, one receives the following representation of the references:
\lstset{literate=
  {á}{{\'a}}1 {é}{{\'e}}1 {í}{{\'i}}1 {ó}{{\'o}}1 {ú}{{\'u}}1
  {Á}{{\'A}}1 {É}{{\'E}}1 {Í}{{\'I}}1 {Ó}{{\'O}}1 {Ú}{{\'U}}1
  {à}{{\`a}}1 {è}{{\`e}}1 {ì}{{\`i}}1 {ò}{{\`o}}1 {ù}{{\`u}}1
  {À}{{\`A}}1 {È}{{\'E}}1 {Ì}{{\`I}}1 {Ò}{{\`O}}1 {Ù}{{\`U}}1
  {ä}{{\"a}}1 {ë}{{\"e}}1 {ï}{{\"i}}1 {ö}{{\"o}}1 {ü}{{\"u}}1
  {Ä}{{\"A}}1 {Ë}{{\"E}}1 {Ï}{{\"I}}1 {Ö}{{\"O}}1 {Ü}{{\"U}}1
  {â}{{\^a}}1 {ê}{{\^e}}1 {î}{{\^i}}1 {ô}{{\^o}}1 {û}{{\^u}}1
  {Â}{{\^A}}1 {Ê}{{\^E}}1 {Î}{{\^I}}1 {Ô}{{\^O}}1 {Û}{{\^U}}1
  {Ã}{{\~A}}1 {ã}{{\~a}}1 {Õ}{{\~O}}1 {õ}{{\~o}}1
  {œ}{{\oe}}1 {Œ}{{\OE}}1 {æ}{{\ae}}1 {Æ}{{\AE}}1 {ß}{{\ss}}1
  {ű}{{\H{u}}}1 {Ű}{{\H{U}}}1 {ő}{{\H{o}}}1 {Ő}{{\H{O}}}1
  {ç}{{\c c}}1 {Ç}{{\c C}}1 {ø}{{\o}}1 {å}{{\r a}}1 {Å}{{\r A}}1
  {€}{{\euro}}1 {£}{{\pounds}}1 {«}{{\guillemotleft}}1
  {»}{{\guillemotright}}1 {ñ}{{\~n}}1 {Ñ}{{\~N}}1 {¿}{{?`}}1
  {·}{{\(\cdot\)}}1  {•}{{\(\bullet\)}}1
}
\begin{lstlisting}[fontadjust=false,breaklines=true,basicstyle=\tiny]
(1] A. RÉNYI, On oonnected g1'aphs, Magyar Tud. Aka.d. Mat. Kutato Iut. Ki)zl., 4, 1959,
p. 385-388.
[2] G. FORD, G. UHLENBECK, Combinatorial problems in the tlteor·y of graphs, Proceedings of
the National Acttdemy of Sciences of the U. S. A., 42, 1956, p. 122-128.
(3] J. \\V. MooN, En!mterating Labeled 1'rees, dans: Graph Theory an•l Theoretica.l Physic\~,
editor F. Httrary, Academie Press, New York-London, 1967, p. 261-271.
\end{lstlisting}

\begin{figure}[tb]
\includegraphics[width=\textwidth]{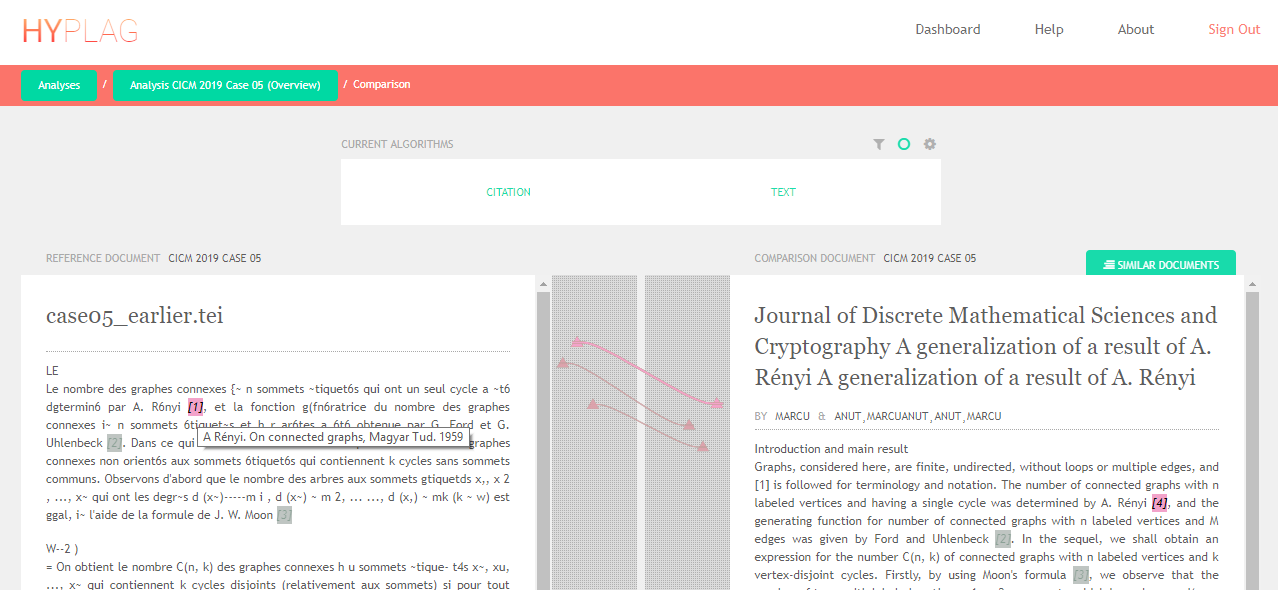}
\caption{Visualization of case 5 in HyPlag prototype}\label{fg.case5}
\end{figure}

The quality of the extracted references was too low to match references and in-text citations in the documents. To demonstrate the potential of the approach, we manually corrected the OCR text. Figure \ref{fg.case5} shows the result: the first three citations appear in identical order in both documents resulting in a high citation-based similarity score.
$e_5$ has been neither retracted nor cited so far.

\paragraph{Case 6 (topical relatedness)} consists of $l_6$. The main result of this paper had been proven by an earlier article $e_6$ using similar methods (or, as \url{https://mathscinet.ams.org/mathscinet-getitem?mr=2177689} states: ``The same theory can be found in an essentially identical presentation in [the] earlier paper [...] not cited by the present author."). A subsequent discussion by expert reviewers remained inconclusive; thus, the paper has not been retracted. Technical PD tools fail to detect any significant similarity of the articles. Perhaps due to the awareness created by the reviewers' discussion, $l_6$ received only two (self-)citations, while $e_6$ is cited frequenlöy.

\paragraph{Case 7 (distribution stopped)} represents the translation of a book. The German original $e_7$ from 2003 was translated to English and published in 2009 as $l_7$.%
The publisher stopped the distribution of the English version\footnote{\url{https://www.emis.de/misc/articles/ext05526289.html}}, which therefore was digitally unavailable to us. Despite its short period of availability, $l_7$ has been cited at least three times, while only one citation of $e_7$ is known so far.

\paragraph{Case 8 (identical)} is an example for a work that reuses an earlier work, changes the mathematical notation, and presents a weaker result. The later paper $l_8$ which reused content from $e_8$ is still online without referencing the original work. The later paper received three citations according to Google scholar, was downloaded 4013 times, and viewed 7615 times according to the publisher\footnote{\url{https://www.scirp.org/journal/PaperInformation.aspx?PaperID=3820}}. Despite the significant overlap, the publisher did not issue a retraction but only a statement of priority for the original result. The statement appeared inconspicuously in a later issue and is not linked from the article itself. The similarity of the text and the references, although slightly below a critical threshold, should have issued at least a warning by a PD system.

\paragraph{Case 9 (unclear)} comprises $l_9$ which adopted content from the earlier work $e_9$. The notation was changed and the text differs (text similarity 0.09). However, expert reviewers qualified the later article as: ``derivative work". So far, no retraction has been issued. According to the publisher\footnote{\url{https://link.springer.com/article/10.1007\%2FBF02463791}}, the derived work was downloaded 27 times and received one citation.

\paragraph{Case 10 (unclear)} consists of the paper $l_{10}$, parts of which reuse material from $e_{10}$ literally identical. However, the overall text similarity is below the critical threshold (score 0.16). Another noticeable difference is that one paper %
uses the computer algebra system Maple while the other paper uses Mathematica. No retraction has been issued. According to a later comment\footnote{\url{https://doi.org/10.1016/j.camwa.2011.01.043}}, the later article derives incorrect formulae. Nevertheless, both articles achieved citation counts that are well above the average for math articles. The citations seem to originate from a rather peculiar community. %
A PD system allowing for a uniform representation of Maple/Mathematica content would have facilitated a clear detection of the similarities. However, such as system is a distant prospect.

\paragraph{Case 11 (compilation of text elements)} consists of paper $l_{11}$ which combines material from $e_{11,1}$ and a paper\footnote{\url{https://math.berkeley.edu/~kwray/papers/string_theory.pdf}} not part of the zbMATH corpus. A reviewer who had critically noted the authors' way of working before indicated the case to zbMATH. So far, the journal has not reacted on the comments in any way. Due to the low visibility of the journal, the compilation seem to have had little impact so far. The comparably low text similarity of the article derives intrinsically from being a compilation of two sources. For such cases, an adapted measure resulting in an adequate warning would be desirable. Humans spot the respective adaptations quickly.

\subsection{Requirements for a plagiarism detection system at zbMATH}
The investigation of the selected cases indicates that the application of a purely text-based system, such as the commercial service of iThenticate, appears insufficient for analyzing content overlap in mathematical publications. Many publishers already use iThenticate as part of their submission pipelines.

The major obstacle for using the open source solution HyPlag is the availability of high quality sources. While PDF files were often available, the mathematical formulae could not be extracted from these PDFs. Therefore, the math-based similarity detection of HyPlag \cite{Meuschke2017} could not be evaluated in this paper.
Moreover, some PDFs are of a low quality ans resulted in OCR text that is too erroneous for citation matching.
Another problem with the PDF sources was that figures could not be identified.

\section{Conclusion and Outlook} \label{sec.concloutl}
We created an openly accessible %
dataset of 149 papers with noticeable content reuse in zbMATH. The dataset can serve as a training set or test set, e.g., for plagiarism detection competitions, such as PAN. In a second step, we extended the 149 confirmed cases of NCR with cases for which the content similarity was eventually rated as legitimate. To not discredit authors who were incorrectly accused of wrongdoing, we refrain from publishing the complete dataset of 446 cases. Instead, we composed a list of 11 typical cases that illustrate the spectrum of reported content reuse. Moreover, we will continue our analysis with more cases to derive general patterns.

Using the 11 cases, we investigated how the plagiarism detection system HyPlag would perform in identifying the documents as suspicious. In a recent study \cite{Meuschke2019}, we applied the system to a large test collection producing only a small number of false positives. However, nine of ten test cases in \cite{Meuschke2019} could have been discovered using traditional text-based detection methods. In contrast, for the zbMATH collection, this number is only three of eleven.
For the text-based and the math-based detection methods, we needed to transcribe the test data manually since the sources were unavailable and the quality of formulae, citations, and figures extracted from PDF was insufficient for reliably matching these features. However, we demonstrated that our detection system would have discovered the similarities in content if the data would be available in LaTeX or XHTML format.

The dataset of 446 cases supports zbMATH's work towards the goal of installing a system that supports the editor in identifying potentially suspicious documents, even if the final decision is not to issue a public note on content reuse. In other words, the notification threshold of the system needs to be lower than for most plagiarism detection systems. Furthermore, the system must enable the zbMATH editor to easily understand why a document has been retrieved as potentially suspicious. This requirement is even more important than automatically performing a highly accurate binary classification of documents as suspicious or unsuspicious. To achieve this goal, visualizing the topical similarity is a key features required of the future system.

The next steps for realizing such a system are to establish an automated workflow for receiving the full-texts of the papers submitted to zbMATH. Moreover, we need to obtain mathematical formulae in a machine-readable format for at least a fraction of the zbMATH collection.  We will continue our efforts to extract \LaTeX formulae from PDF documents and are looking forward to the results of this years CHROME competition \footnote{\url{https://www.cs.rit.edu/~crohme2019/index.html}}. Especially the results for machine-readable formulae will be the foundation to conceive more sophisticated math-based detection methods.
In the long run, we plan to lower the notification threshold for content reuse, which will undoubtedly require more sophisticated detection methods for formula similarity. The idea is to also identify papers that did not plagiarize but have limited novelty.

\paragraph*{Acknowledgements}
This work was supported by the German Research Foundation (DFG grant GI-1259-1). We thank Alexandar Perovi\'c for identifing that \zbl{1360.47003} was a false positive match by the matching heuristic presented in the first ersion of this publication.

\printbibliography[keyword=primary]
\appendix
\section{List of documents with noticeable content reuse}\label{sec.list}
\scriptsize
\zbl{1360.53021},
\zbl{1357.30013},
\zbl{1353.39029},
\zbl{1353.30019},
\zbl{1345.15011},
\zbl{1359.62073},
\zbl{1356.01026},
\zbl{1337.16003},
\zbl{1354.47018},
\zbl{1340.90030},
\zbl{1345.92082},
\zbl{1318.46035},
\zbl{1400.34041},
\zbl{1388.42037},
\zbl{1388.42036},
\zbl{1343.65150},
\zbl{1330.35490},
\zbl{1322.93076},
\zbl{1321.81036},
\zbl{1307.65177},
\zbl{1308.81133},
\zbl{1358.47017},
\zbl{1309.65163},
\zbl{1304.57008},
\zbl{1325.47059},
\zbl{1301.16002},
\zbl{1293.65167},
\zbl{1359.62055},
\zbl{1291.30077},
\zbl{1323.65125},
\zbl{1328.47074},
\zbl{1328.47073},
\zbl{1299.65168},
\zbl{1295.35151},
\zbl{1294.35189},
\zbl{1290.26023},
\zbl{1282.91334},
\zbl{1279.91096},
\zbl{1281.35058},
\zbl{1306.90186},
\zbl{1311.90164},
\zbl{1273.91086},
\zbl{1287.81012},
\zbl{1386.18011},
\zbl{1301.45006},
\zbl{1290.18001},
\zbl{1278.68235},
\zbl{1278.53033},
\zbl{1271.54029},
\zbl{1271.39024},
\zbl{1266.65214},
\zbl{1266.33002},
\zbl{1264.34048},
\zbl{1342.34118},
\zbl{1266.30001},
\zbl{1265.39016},
\zbl{1264.81239},
\zbl{1257.11089},
\zbl{1250.78038},
\zbl{1246.90035},
\zbl{1246.90034},
\zbl{1250.78039},
\zbl{1252.68177},
\zbl{1234.34034},
\zbl{1364.47004},
\zbl{1399.35153},
\zbl{1274.76184},
\zbl{1252.83109},
\zbl{1288.49015},
\zbl{1249.60023},
\zbl{1250.47059},
\zbl{1231.83033},
\zbl{1227.34015},
\zbl{1219.30004},
\zbl{1236.58009},
\zbl{1230.46033},
\zbl{1213.60020},
\zbl{1211.34093},
\zbl{1211.34092},
\zbl{1295.91090},
\zbl{1242.49079},
\zbl{1234.60021},
\zbl{1262.11083},
\zbl{1221.81113},
\zbl{1234.60020},
\zbl{1211.46021},
\zbl{1217.34137},
\zbl{1211.11127},
\zbl{1203.06007},
\zbl{1203.06006},
\zbl{1212.49026},
\zbl{1193.35074},
\zbl{1191.35223},
\zbl{1253.60034},
\zbl{1235.37020},
\zbl{1186.54007},
\zbl{1189.35123},
\zbl{1188.16002},
\zbl{1183.37156},
\zbl{1371.91006},
\zbl{1371.91005},
\zbl{1258.74210},
\zbl{1257.78018},
\zbl{1192.34093},
\zbl{1195.55004},
\zbl{1212.60016},
\zbl{1201.60017},
\zbl{1184.20030},
\zbl{1176.91147},
\zbl{1173.90327},
\zbl{1206.34097},
\zbl{1177.35217},
\zbl{1170.34353},
\zbl{1173.34354},
\zbl{1279.90096},
\zbl{1153.91544},
\zbl{1189.35124},
\zbl{1177.35218},
\zbl{1175.86006},
\zbl{1162.30319},
\zbl{1170.42304},
\zbl{1165.35336},
\zbl{1162.30309},
\zbl{1153.86318},
\zbl{1154.94319},
\zbl{1250.49003},
\zbl{1166.47308},
\zbl{1153.91523},
\zbl{1155.26016},
\zbl{1157.05036},
\zbl{1162.83357},
\zbl{1139.81335},
\zbl{1213.35364},
\zbl{1169.46304},
\zbl{1169.42310},
\zbl{1144.81475},
\zbl{1141.90010},
\zbl{1231.93121},
\zbl{1132.14304},
\zbl{1144.46044},
\zbl{1134.60382},
\zbl{1129.83326},
\zbl{06921286}.

\end{document}